\documentclass[10pt]{dis03}
\usepackage{epsfig,graphics,amsmath}

\begin{document}

\title{Measurement of inclusive jet cross-sections at low $Q^2$ at HERA\footnote{
to be published in: {\it Proceedings of DIS 2003,
XI International Workshop on Deep Inelastic Scattering,
St. Petersburg, 23-27 April 2003}}}

\author{
  Arnd E. Specka\\
  \it\small 
        Laboratoire Leprince-Ringuet, Ecole Polytechnique, F-91128 Palaiseau, France\\
        present address: DESY, Notkestr. 85, D-22607 Hamburg, Germany\\
        e-mail: specka@poly.in2p3.fr}
\maketitle

\begin{abstract}
\noindent
The single-inclusive cross-section for jet production in
deep inelastic e-p scattering (DIS) at HERA is measured for
photon virtualities $Q^2$ between 5 and 100~GeV$^2$,
differentially in $Q^2$, in transverse jet energy $E_T$, in
$E_T^2/Q^2$ and in pseudorapidity $\eta_\mathrm{LAB}$. In most of the
phase space these data are well described by QCD
calculations in next-to-leading order (NLO) using a
renormalization scale $\mu_R=E_T$.  Significant
discrepancies are observed only for jets in the proton beam
direction with $E_T$ below 20~GeV and $Q^2$ below
20~GeV$^2$.
\end{abstract}
\medskip

%
%\section{Introduction}
%
Jet production in DIS is an ideal testing ground for
perturbative QCD (pQCD). Experimental results on (multi) jet
production at HERA and elsewhere cover a wide range of $Q^2$
and have been found to be well described by pQCD in
next-to-leading order in the strong coupling constant
$\alpha_S$. The study of inclusive jet production, in
particular, gives access to a bigger phase space, and avoids
phase space regions where pQCD calculations are infrared
sensitive. The data presented here close the gap of
inclusive jet production at $Q^2$ below 100~GeV$^2$. For
details on the experimental measurement and on the theory
predictions see~\cite{ref:IncJets} and references therein.
\vspace{-10pt}

\paragraph*{Event Selection and Measured Observables\\}
The present analysis is based on a total integrated
luminosity of 21.1~pb$^{-1}$ of e$^+$-p collision data
collected with the H1 detector~\cite{ref:H1Det} in 96 and
97. DIS events are selected by requiring the presence in the
electromagnetic calorimeter of an electron with $E>10\,$GeV
and with an angle above 156$^{\mathrm{o}}$ with respect to the
proton beam direction.  The kinematic range is then defined
by requiring a virtuality $Q^2$ between 5 and 100~GeV$^2$
and an inelasticity $y$ between 0.2 and 0.6, as determined
from the scattered electron.
 
The jet search is performed in the Breit frame where the
partons bounce off the virtual photon like from a ``brick
wall'', if one assumes the naive quark parton model.  In
this frame --- as in all reference systems where the
$\gamma^*$ and the proton collide head-on --- the transverse
parton energy can be considered to stem mainly from QCD
radiation.

Jets are searched with the longitudinally invariant,
inclusive $k_t$-algorithm, a clustering
algorithm with a transverse-energy-weighted distance measure
in the $\eta$-$\phi$-plane. This algorithm has been
demonstrated to be stable against infrared and collinear
divergencies and has been extensively used in previous jet
analyses at HERA.

For the inclusive cross-section all jets with transverse
energy $E_T$ in the Breit-frame above 5~GeV are considered.
Furthermore, the pseudo-rapidity $\eta_\mathrm{LAB}$ of the
jets (when Lorentz-transformed back to the laboratory frame)
are required to lie between -1 and 2.8, where
electromagnetic and hadronic energy measurement is
excellent.

Two sources dominate the systematic error of the measured
cross-sections. First, the two Monte-Carlo models used to
correct the data for detector effects and QED
radiation, RAPGAP and DJANGO/ARIADNE, differ by typically
5--10\% in cross-section. Second, a 3\% uncertainty in the
energy scale of the hadronic liquid argon calorimeter leads
to a typical error on the cross-sections of 10--15\%.
\vspace{-10pt}

\paragraph*{Perturbative QCD predictions in NLO\\}
The measured jet cross-sections are compared to pQCD
predicitions at first (LO) and second (NLO) order in
$\alpha_S$. These calculations are performed using the
DISENT program~\cite{ref:DISENT} and CTEQ5M/CTEQ5L parton
distribution functions. Note that these calculations only
take into account the direct photon contribution and neglect
any virtual photon structure. The cross-sections have been
found to be rather insensitive to even large variations of
the factorization scale for which the momentum transfer $Q$
has been used throughout.

Because the perturbation series is truncated, fixed order
QCD calculations exhibit an explicit dependence on the
renormalization scale $\mu_R$. Among the two possible hard
scales in DIS jet events, $Q$ and jet transverse energy
$E_T$, the latter has been mainly chosen for $\mu_R$, but
the alternate choice $\mu_R=Q$ has been studied as well (see
below).  In order to assess the sensitivity of the theory
prediction to the value of the renormalization scale,
$\mu_R$ has been raised (lowered) by a factor of 2 (0.5).

\begin{figure}[tbh]
      \includegraphics[width=.95\linewidth]{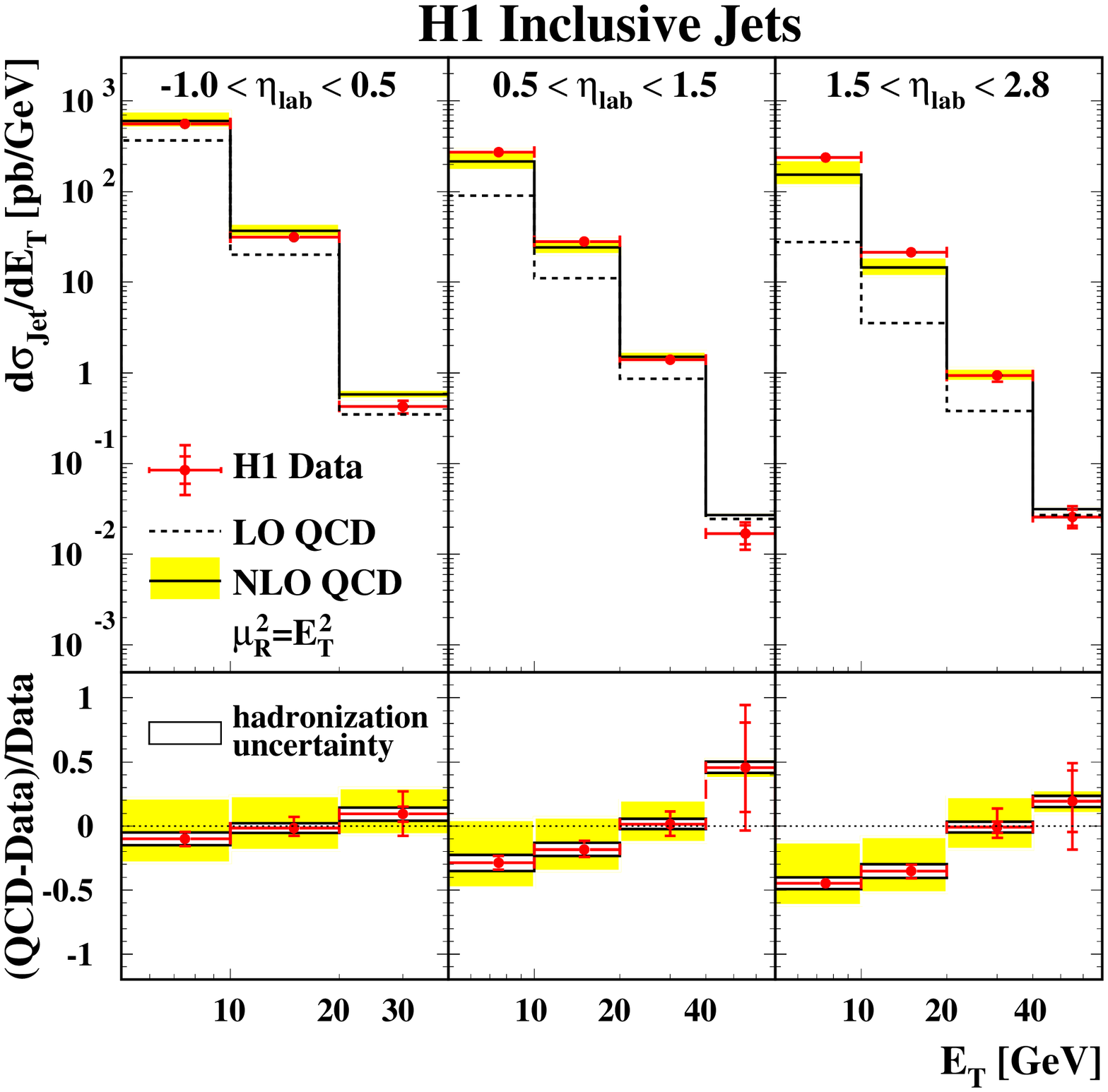}
      \vspace*{-5mm}
      \caption{\it $E_T$ dependence in backward, central, and
        forward regions 
    \label{fig:dsdetalleta}
    }
\end{figure}

Furthermore, for a detailed comparison of the parton-level
QCD calculations to the measured data a hadronization
correction has been applied to the former. The Lund string
model of hadronization together with two different models of
soft gluon emission (CDM and ME+PS) are used.  In the
kinematic range under study here, hadronization is found to
decrease the cross-section typically by 5--15\%.
\vspace{-10pt}

\paragraph*{Experimental Results\\}
\begin{figure}[tbh]
      \includegraphics[width=.95\linewidth]{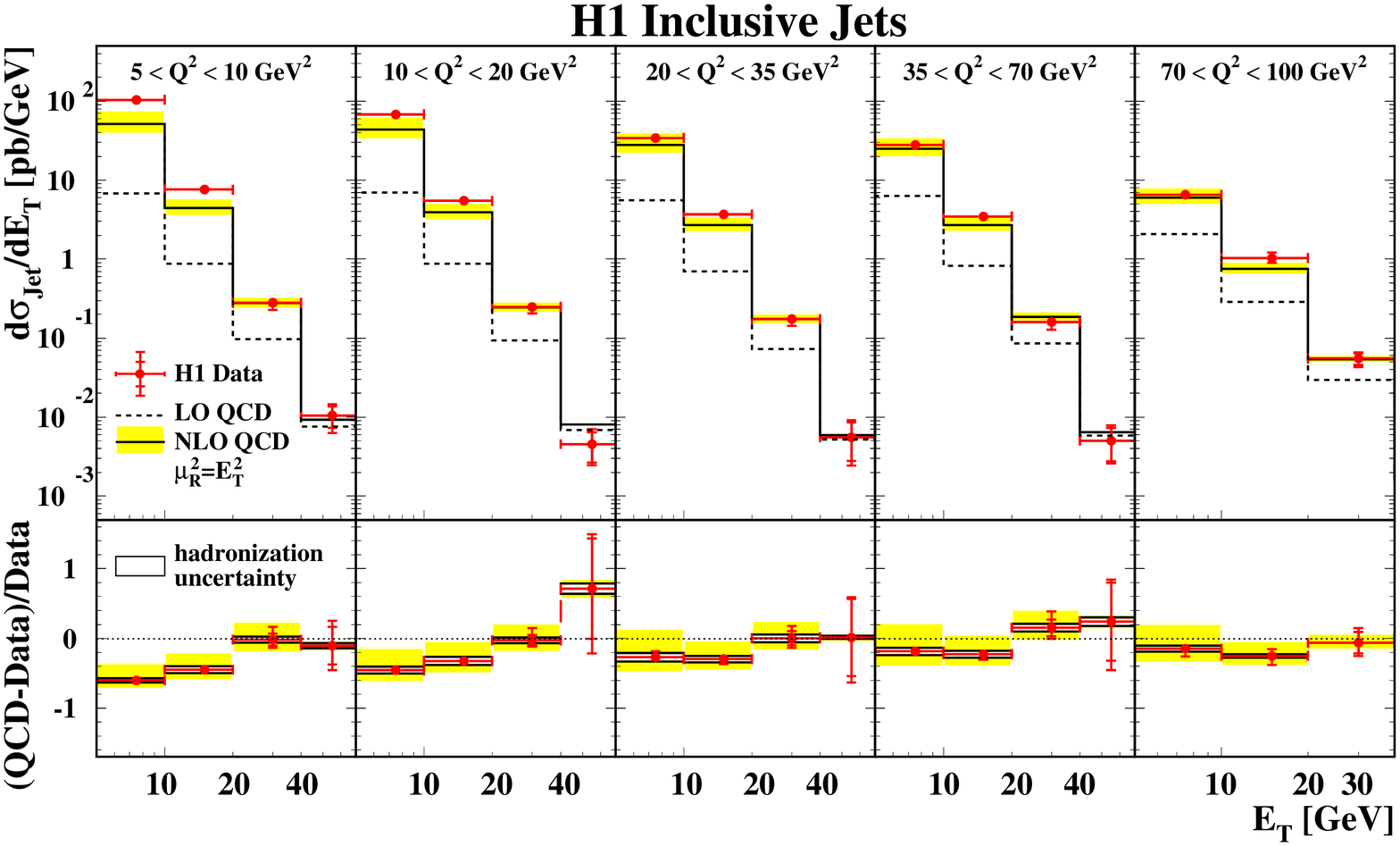}
      \vspace*{-5mm}

      \caption{\it $Q^2$ dependence $\mathrm{d}\sigma_\mathrm{Jet}/\mathrm{d} E_T(E_T)$ in the forward region.
    \label{fig:dsdetforward}
    }
\end{figure}
In figure~\ref{fig:dsdetalleta} the $E_T$-dependence of the
inclusive jet cross-section is compared to QCD predictions
in three ranges of pseudorapidity, here called backward
($-1.0<\eta_\mathrm{LAB}<0.5$), central
($0.5<\eta_\mathrm{LAB}<1.5$), and forward
($1.5<\eta_\mathrm{LAB}<2.8$). In the upper part of these
(and all following plots) the data are compared with LO and
NLO QCD calculations at parton level, whereas in the lower
part the relative difference between data and NLO QCD
prediction after applying the hadronization correction is
shown. The uncertainty of the NLO QCD prediction due to the
variation of $\mu_R$ is indicated by a shaded band.

The remarkably good description of the experimental data by
NLO QCD in the backward and central region is due to the
substantial NLO/LO corrections. In the forward region where
these are highest, however, the QCD prediction falls short
of the data by about 40\% for $E_T$ below 20~GeV.

Figure~\ref{fig:dsdetforward} takes a closer look at the
$Q^2$ dependence of
$\mathrm{d}\sigma_\mathrm{Jet}/\mathrm{d} E_T$ in the
forward region.  For $Q^2$ above 20~GeV$^2$ the data are
well described by the QCD prediction, again owing to the
large NLO/LO corrections which reach up to a factor of 4.
For $Q^2 < 20\,$GeV$^2$, in contrast, the theoretical
prediction falls short of the data by up to 50\% for $E_T$
below 20~GeV where NLO/LO corrections are largest.

\begin{figure}[tbh]                                                               
  \raisebox{-6mm}{
    \parbox[b]{.49\linewidth}{
      \centering\mbox{
        \includegraphics[width=.95\linewidth]{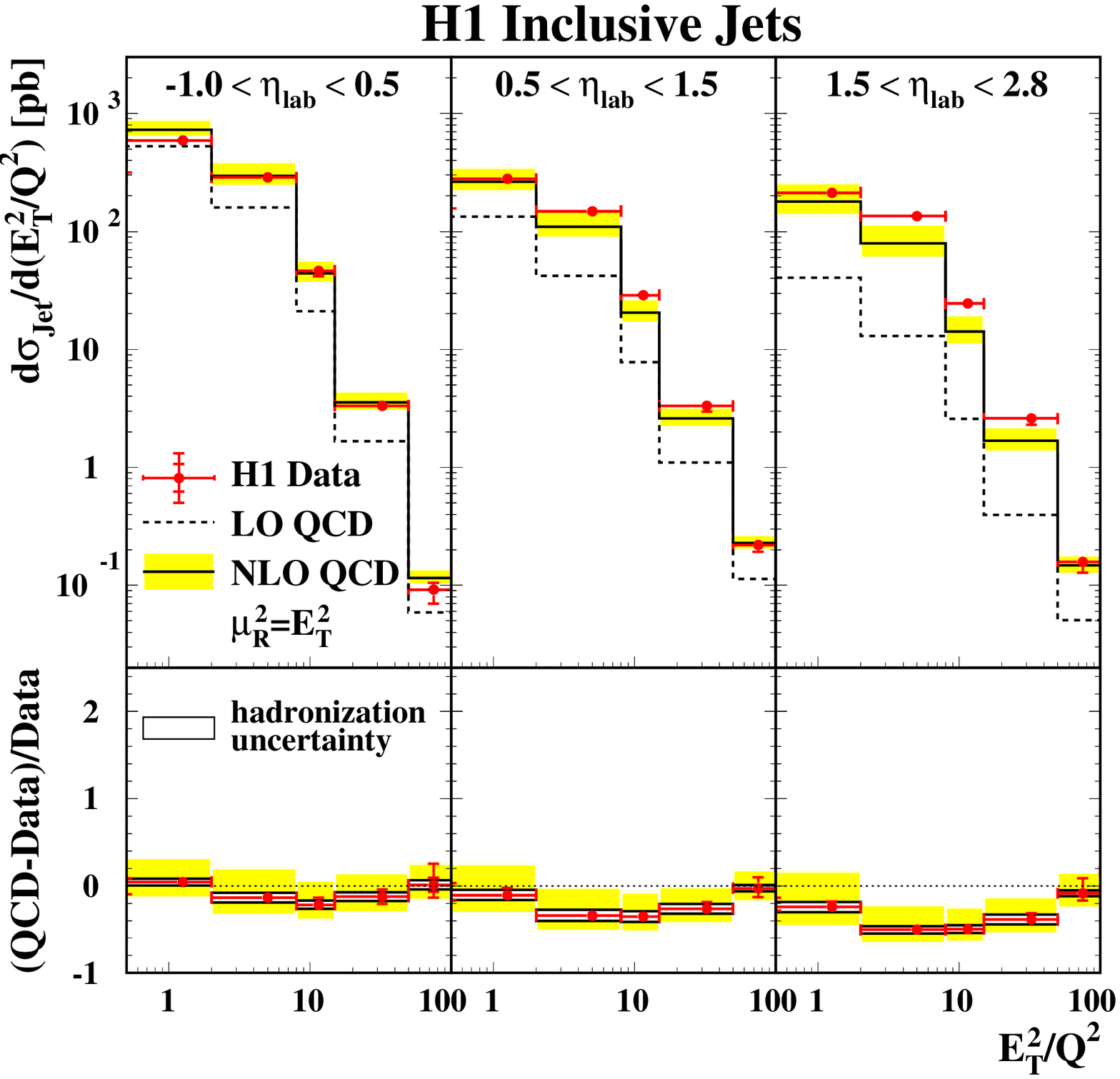}
        }
      }
    }\hfill\vspace{-6mm}\raisebox{-6mm}{
    \parbox[b]{.49\linewidth}{
      \centering\mbox{
        \includegraphics[width=.95\linewidth]{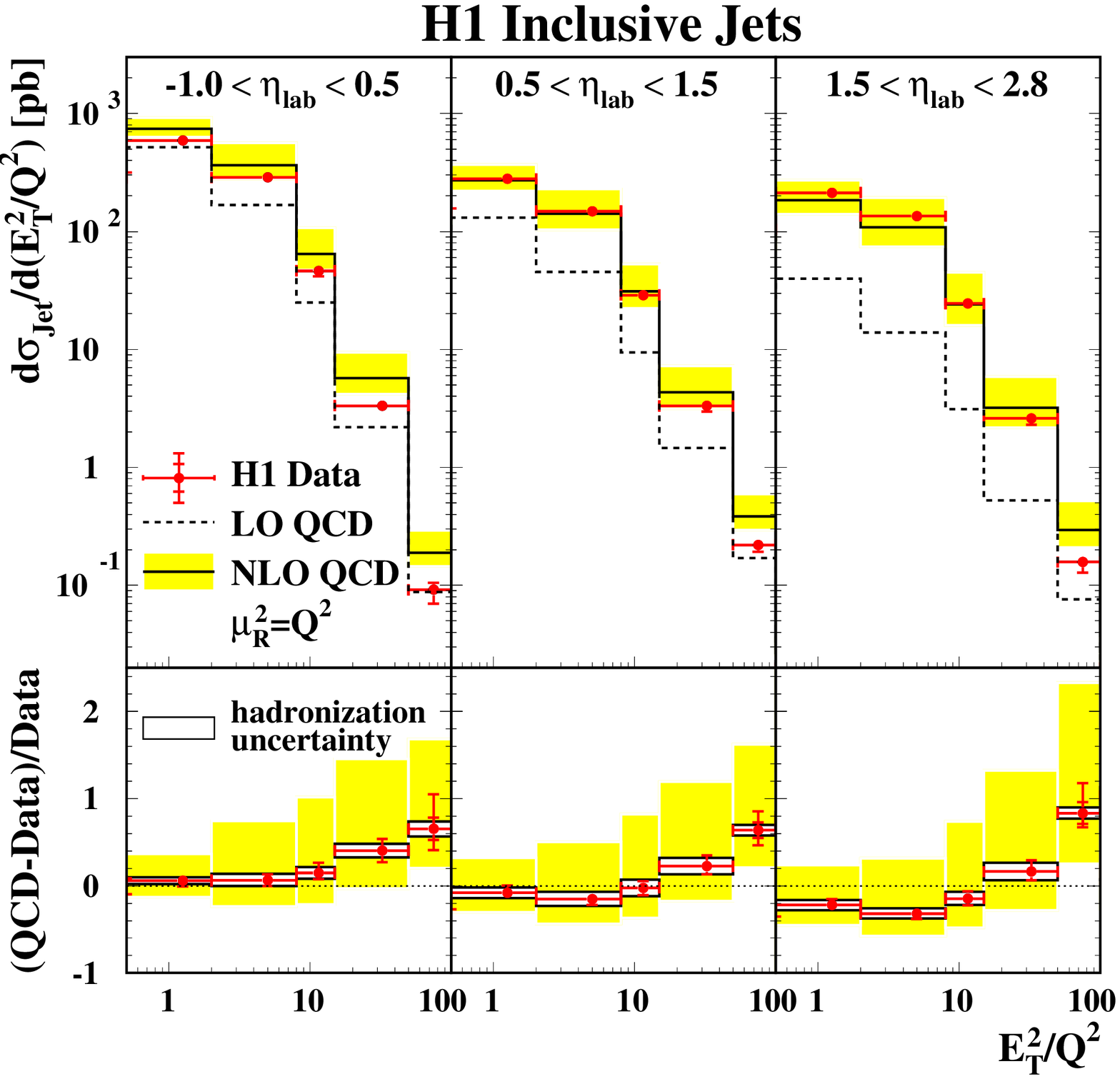}
        }
      }
    }
  \parbox[t]{.49\linewidth}{
    \centering\parbox[t]{.99\linewidth}{            
      \rule{0pt}{1em}\vspace{-\baselineskip}\vspace{-7pt}\newline 
      \caption{\it$\mu_R=E_T$\label{fig:dsdrscuret}}
      }
    }                
  \hfill                                                                          
  \parbox[t]{.49\linewidth}{
    \centering\parbox[t]{.99\linewidth}{            
      \rule{0pt}{1em}\vspace{-\baselineskip}\vspace{-7pt}\newline 
      \caption{\it$\mu_R=Q$\label{fig:dsdrscurq2}}
      }
    }                
  \parbox[t]{\linewidth}{
    \centering\parbox[t]{.99\linewidth}{            
      \rule{0pt}{1em}\vspace{-\baselineskip}\vspace{2pt}\newline 
      {\it Inclusive jet cross-section measured differentially in the scale ratio $E_T^2/Q^2$\\
        and comparison to pQCD calculations with two alternative choices of $\mu_R$}
      }
    }                
\end{figure}                                                                    
In order to study the interplay of the two possible scales
in DIS jet production, figure~\ref{fig:dsdrscuret} shows the
inclusive jet cross-section differentially in the scale
ratio $E_T^2/Q^2$, in the three $\eta_\mathrm{LAB}$ ranges.
Only the backward region is well described by NLO QCD,
whereas in the central and forward region the theory
prediction lies 30--50\% below the measured data for
$2<E_T^2/Q^2<50\,$GeV$^2$. This medium range of the scale
ratio is dominated by small values of $E_T^2$ and $Q^2$.

So far, $E_T$ has been used as the renormalization scale in
the NLO calculation. Switching to $\mu_R=Q$ changes the
situation as can be seen from figure~\ref{fig:dsdrscurq2}.
Here, the strongest discrepancies occur for very high values
of the scale ratio, i.e. where $Q^2$ is small, irrespective
of $\eta_\mathrm{LAB}$. This suggests that $Q^2$ is not an
appropriate choice for $\mu_R$ when $E_T^2$ is much higher
than $Q^2$. It should be noted that this choice of $\mu_R$
strongly increases the scale uncertainties, making the QCD
calculations less predictive.  \vspace{-10pt}

\paragraph*{Summary\\}
The single-inclusive jet cross-section in DIS has been
measured double differentially for $Q^2$ below 100~GeV$^2$
with the H1 detector at HERA.  With $E_T$ as the
renormalization scale, the data are well described by NLO
QCD calculations except when both $Q^2$ and $E_T$ are
relatively small. The correlation of large NLO/LO
corrections and high sensitivity to $\mu_R$ variations with
poor agreement between data and theory strongly suggests that
the inclusion of higher order (e.g. NNLO) terms in the QCD
calculations is necessary in order to describe the data.


\vspace{-10pt}

\begin{thebibliography}{10}\vspace{-10pt}
\bibitem{ref:IncJets}
  C. Adloff {\it et al.}, Phys. Lett. {\bf B542} (2002) 193
\vspace{-6pt}
\bibitem{ref:H1Det}
  I.~Abt {\it et~al.}, 
  Nucl. Instrum. Meth. {\bf A386} (1997) 310
\vspace{-6pt}
\bibitem{ref:DISENT} 
  S. Catani and M. H. Seymour, Nucl. Phys.
  {\bf B 485} (1997) 291
\end{thebibliography}
\end{document}